\documentclass[11pt,amssymb,nofootinbib,showkeys]{revtex4}
\usepackage[latin9]{inputenc}
\usepackage{textcomp}
\usepackage{amsmath}
\usepackage{amssymb}
\usepackage{esint}
\usepackage{graphicx}
\usepackage{epstopdf}
\usepackage[english]{babel}

\makeatletter
\@ifundefined{textcolor}{}
{%
 \definecolor{BLACK}{gray}{0}
 \definecolor{WHITE}{gray}{1}
 \definecolor{RED}{rgb}{1,0,0}
 \definecolor{GREEN}{rgb}{0,1,0}
 \definecolor{BLUE}{rgb}{0,0,1}
 \definecolor{CYAN}{cmyk}{1,0,0,0}
 \definecolor{MAGENTA}{cmyk}{0,1,0,0}
 \definecolor{YELLOW}{cmyk}{0,0,1,0}
}

\usepackage{amsthm}\usepackage{amscd}\usepackage{bm}\usepackage{bbm}

\let\vec\boldvec%

\makeatother

\begin{document}

\title{The emission of electromagnetic radiation from a quantum system interacting with an external noise: A general result}


\author{Sandro Donadi}

\email{sandro.donadi@ts.infn.it}

\affiliation{Department of Physics, University of Trieste, Strada Costiera 11,
34151 Trieste, Italy}

\affiliation{Istituto Nazionale di Fisica Nucleare, Trieste Section, Via Valerio
2, 34127 Trieste, Italy}

\author{Angelo Bassi}

\email{bassi@ts.infn.it}

\affiliation{Department of Physics, University of Trieste, Strada Costiera 11,
34151 Trieste, Italy}

\affiliation{Istituto Nazionale di Fisica Nucleare, Trieste Section, Via Valerio
2, 34127 Trieste, Italy}


\begin{abstract}
We compute the spectrum of emitted radiation by a generic quantum system interacting with an external classic noise. Our motivation is to understand this phenomenon within the framework of collapse models. However the computation is general and applies practically to any situation where a quantum system interacts with a noise. The computation is carried out at a perturbative level. This poses problems concerning the correct way of performing the analysis, as repeatedly discussed in the literature. We will clarify also this issue.   
\end{abstract}
\maketitle

\section{Introduction}
In this paper we analyze the dynamics of a non-relativistic charged quantum system under the influence of an external classical random scalar field. We focus on the spectrum of the radiation emitted by the system, due to the interaction with the field. 

The study of the radiation emission is particularly important in the context of collapse models, because up to now this process sets the strongest bound on the possible values of the phenomenological parameters defining these models can take~\cite{ap,sci}. The idea is the following: collapse models modify the standard quantum evolution given by the Schr\"odinger equation, by adding a nonlinear interaction with a random scalar field, which induces the collapse of the wave function~\cite{grw,csl,rep1,rep2}. In the case of a charged particle, the random field forces the particle to emit radiation, whereas standard quantum theory predicts no emission (if the particle is free). Therefore this situation represents a case-study for testing collapse models against the standard theory~\cite{mm0,mm1,mm2,mm3,mm4}. 

This calculation finds application also in the theory of open quantum systems. A charged particle in a bath, typically interacting with the other particles via a position dependent force, would also emit radiation.

The problem of radiation emission was already studied in the literature of collapse models. In~\cite{fu3} a calculation for the free particle was carried out to the first perturbative order using the Continuous Spontaneous Localization (CSL) model~\cite{csl}. This result was confirmed and generalized to the case of a non white noise in~\cite{ar}, where the analysis was also extended to the case of an hydrogenic atom. Later in~\cite{bd} the formula for the emission rate was computed by using the simpler Quantum Mechanics with Universal Position Localizations (QMUPL) model~\cite{qmupl1,qmupl2} for the case of a free particle and a harmonic oscillator. However, the formula found in~\cite{bd} did not agree with the one found in ~\cite{fu3,ar} as they should have: in the case of a white noise, the emission rate from a free particle found in~\cite{bd} turned out to be twice that of~\cite{fu3,ar}. The origin of this discrepancy was clarified in~\cite{abd}, where the perturbative calculation in the CSL model was repeated. The result, in the white noise case, was found to be in agreement with the one of~\cite{bd}. However it was also shown that, when the computation is generalized to the non white noise case, an unphysical contribution appears. In fact the formula for the rate was found to be~\cite{abd}:
\begin{equation}
\frac{d\Gamma}{dk}\;=\;\frac{\lambda\hbar e^{2}}{4\pi^{2}\varepsilon_{0}c^{3}m_{0}^{2}r_{C}^{2}k}\;\left[\tilde{f}(0)+\tilde{f}(\omega_{k})\right].\label{eq:rate non white}
\end{equation}
where $\hbar$, $c$ and $\varepsilon_{0}$ have the usual meaning, $\lambda$ and $r_{C}$ are two parameters characterizing the CSL model~\cite{csl}, $m_0$ the mass of a nucleon, $k=|{\mathbf k}|$ with $\mathbf k$ the photon wave vector, and 
\begin{equation}\label{ftilde}
\tilde{f}(\omega):=\int_{-\infty}^{+\infty}f(s)e^{i \omega s}ds,
\end{equation}
is the noise spectral density where $f(s)$ is the noise time correlation function. In the white noise case $f(s)=\delta(s)$, which implies $\tilde{f}(\omega)=1$ for any $\omega$. The second term in Eq.~(\ref{eq:rate non white}) tells that the probability of emitting a photon with wave vector $k$ is proportional to the spectral density of the noise at the frequency $\omega_{k}=kc$. This is an expected contribution. On the other hand, the first term is proportional to the spectral density of the noise at zero energy and is physically suspicious. In fact, the zero energy component of the noise is not  expected to contribute to the emission of photons with an arbitrarily high energy. More precisely, as we will show, this energy non conserving term is exactly what in standard perturbation theory are known as ``non-resonant terms"~\cite{Sakurai2}. In general, these terms should not give any important contribution when computing the emission rate. This is not the case of Eq.~(\ref{eq:rate non white}).
This shows that there are problems when standard perturbative techniques are used to find the emission rate. 
A first way out of the problem was found in~\cite{abd}: it was shown that when the computation is repeated by taking wave packets as final states and by confining the noise, then the unphysical contribution proportional to $\tilde{f}(0)$ is not present anymore. 
However, even if this prescription leads to a satisfactory result, it does not really clarify the origin of the unphysical terms. 

A deeper insight to the problem was obtained using the QMUPL model, where an exact treatment of the problem is possible~\cite{basdon}. It was shown that, in the case of a free particle, the unphysical term proportional to $\tilde{f}(0)$ is still present. However, for an harmonic oscillator with frequency $\omega_0$, the unphysical term is suppressed by an exponential damping factor $e^{-\Lambda t}$ with $\Lambda=\frac{\omega_0^2 \beta}{2m}$ and $\beta = \frac{e^2}{6 \pi \epsilon_0 c^3}$. It is important to note that treating the electromagnetic interaction at the lowest order is equivalent to setting $\beta=0$, meaning $\Lambda =0$, in which case the unphysical term $\tilde{f}(0)$ is not suppressed anymore. The same problem arises when $\omega_0=0$, which is the free particle case. 
Therefore, the analysis done in~\cite{basdon} proved that in order to get a physically meaningful result, first the particle cannot be treated as completely free and, second, the electromagnetic interaction cannot be treated at the lowest perturbative order.  
Using the above result, the emission rate for an harmonic oscillator within the CSL model was computed in~\cite{dirk}. The interaction with the noise was treated perturbatively and the one with the electromagnetic field exactly. It was found that, in the free particle limit, the emission rate is equal to the one given in Eq.~\eqref{eq:rate non white} without the unphysical term. 
However, this analysis lacks of generality: the calculations required to solve the Heisenberg equations, while treating exactly the electromagnetic interaction, and this can be done only for simple systems.

Aim of this work is to derive a very general result, which can be applied to a large variety of systems and interactions. Instead of considering a specific model, as done in the works previously cited, we derive a general result for a generic interaction with an external classic noise of the type described in Eq.~\eqref{eq:xi=i}. Moreover, our result will holds for any generic bounded system, contrary to the previous analysis where only the free particle, harmonic oscillator and hydrogenic atom cases were considered. Since we are considering generic systems, the calculation of the emission rate is too complicated to be done exactly, so we need to resort to perturbation theory. According to the results found in~\cite{basdon,dirk}, in order to avoid the presence of unphysical contributions, we have to find a way of including the effect of the higher order electromagnetic contributions to the interaction. A first order analysis would once again lead to the problem one encounters with Eq.~\eqref{eq:rate non white}. A first attempt might be to consider all the diagrams at the next relevant order. However, this would require too long a calculation, since the number of such diagrams is very large (of order of seventy). As we will show, there is a more clever way to take into account the effects of the relevant higher order contributions. The key point is the observation that, as mentioned before, the unphysical term is exactly what in standard perturbation theory is known as ``non resonant" term~\cite{Sakurai2}. The most general and elegant way for avoiding the presence of these terms is to take into account the decay of the propagator. In fact, because of the electromagnetic interaction, the propagator is not stable and can decay\footnote{In principle there is also a similar effect due to the noise, but here we are not interested in computing it.}~\cite{GreinerSPECIAL,Sakurai2}. We will show that, when this effect is taken into account, the unphysical term is not present anymore. As a result, we will be able to find a formula for the emission rate from a generic system. The result applies to all known collapse models and, as we mentioned before, also to open quantum system, where the effect of the environment is modeled by the interaction with a random potential.
Moreover, as mentioned before, this result is also valid for any quantum system interacting with an external classic noise field.

\section{The model}
The starting point is the Schr{\"o}dinger equation:
\begin{equation}\label{eq:xi=i}
i \hbar \frac{d|\psi_{t}\rangle}{dt} = H_{\text{\tiny TOT}} |\psi_{t}\rangle
\end{equation}
with
\begin{equation}\label{Htotbis}
H_{\text{\tiny TOT}}:=H-\hbar \sqrt{\gamma}
\sum_{\ell} N_{\ell} \, w_{\ell,\,t}\,,
\end{equation}
where $H$ is the standard Hamiltonian of the system, $N_\ell$ are set of commuting self-adjoint operators, $\gamma$ is a coupling constant and $w_{\ell,\,t}$ are a set of independent noises such that:
\begin{equation}\label{eq:noise}
\mathbb{E}\left[w_{\ell,\,t}\right]=0\;\;\;\; \textrm{and} \;\;\;\;\mathbb{E}\left[w_{\ell,\,t}w_{\ell',\,t'}\right]=\delta_{\ell\ell'}f(t-t'),
\end{equation}
with $\mathbb{E}$ denoting the average over the noise and $f$ a generic correlation function. In many cases, the index ``$\ell$" is replaced by the coordinate ``$\mathbf{x}$" and the set of noises become a random classic scalar field in space (and time)~\cite{csl,rep1,rep2}\footnote{In principle one could also consider different forms of interaction between the system and the noises. We focus on the one given in Eq.~\eqref{Htotbis} because this is the coupling between the noises and the system taken in the collapse models, which are the models we are mainly interested in studying.}.

We will consider a generic system composed of $N_p$ charged particles. Since the number of particles is fixed, they can be described by using the first quantization formalism. On the contrary, the electromagnetic field will be described by using the second quantization formalism. 

The standard Hamiltonian $H$ contains three terms:
\begin{equation}\label{Hstandard}
H=H_{\text{\tiny P}}+H_{\text{\tiny R}}+H_{\text{\tiny INT}}.
\end{equation}
The first term is the Hamiltonian of the particles, which has the form:
\begin{equation}\label{H0bis}
H_{\text{\tiny P}}=\sum_{j=1}^{N_p}\left(\frac{\mathbf{p}^{2}_j}{2m_j}+ V(\mathbf{x}_j)+ \sum_{i<j=1}^{N_p}U(\mathbf{x}_j-\mathbf{x}_i)\right)   
\end{equation}
with $m_j$ mass of the $j$-th particle of the system, $V(\mathbf{x})$ an external potential and $U(\mathbf{x}_j-\mathbf{x}_i)$ the interaction potential between the $j$-th and the $i$-th particle. We are not making any assumption on the form of the potentials $V$ and $U$, so they are generic\footnote{To be more precise, in principle there is a restriction on $V$ and $U$. As we will see, in the derivation of the emission rate (Eq.~\eqref{final}) a fundamental role is played by the fact that the eigenstates of $H_p$ are not stable and decay because of vacuum fluctuations of the electromagnetic field. This is true only when the imaginary part of the energy shift $\Delta E_i$ defined in Eq.~\eqref{deltaim1} is different from zero. As one can easily realize by inspecting Eq.~\eqref{deltaim1}, $\Delta E_i$ vanishes only in specific and rather pathological cases, which do not apply to typical bounded systems.}.  The second term is the free Hamiltonian of the electromagnetic field (we are working in Coulomb gauge):
\begin{equation}\label{HRbis}
{H}_{\text{\tiny R}}=\int d\mathbf{x}\, \frac{1}{2}\left(\varepsilon_{0}\mathbf{E}_{\perp}^{2}(\mathbf{x})+\frac{\mathbf{B}^{2}(\mathbf{x})}{\mu_{0}}\right)
\end{equation}
where $\varepsilon_{0}$ and $\mu_{0}$ are, respectively, the vacuum permittivity and permeability and $\mathbf{E}_{\perp}=-\frac{\partial\mathbf{A}}{\partial t}$ is the transverse part of $\mathbf{E}$. The last term describes the interaction between the electromagnetic field and the particles:
\begin{equation}\label{HINTbis}
H_{\text{\tiny INT}} = \sum_{j=1}^{N_p}\left(-\frac{e_j}{m_j}\right) \mathbf{A}(\mathbf{x}_j)\cdot\mathbf{p}_j+\sum_{j=1}^{N_p}\frac{e_{j}^{2}}{2m_j}\mathbf{A}^{2}(\mathbf{x}_j).
\end{equation}
Here $e_j$ is the charge of the $j$-th particle of the system and $\mathbf{A(\mathbf{x})}$ is the vector potential which can be expanded in plane waves as: 
\begin{equation}\label{AAA}
{\bf A}({\bf x}) \; = \; \int d{\bf k}\sum_{\lambda} \alpha_k
\left[ \vec{\epsilon}_{{\bf k}, \lambda}\, a_{{\bf k}, \lambda}\, e^{i{\bf k \cdot
x}} + \vec{\epsilon}_{{\bf k}, \lambda}^{*}\, a_{{\bf k}, \lambda}^{\dagger}\, e^{-i{\bf k
\cdot x}} \right],
\end{equation}
with $\alpha_{k}=\sqrt{\hbar/2\varepsilon_{0}\omega_{k}(2\pi)^{3}}$, $\omega_{k}=kc$, $\vec{\epsilon}_{{\bf k}, \lambda}$ the polarization vectors and $a_{{\bf k}, \lambda}$ and $a_{{\bf k}, \lambda}^{\dagger}$, respectively, the annihilation and creation operators of a photon with wave length $\bf k$ and polarization $\lambda$.

Here we are assuming that the evolution is unitary and driven by the total Hamiltonian $H$ in Eq.~\eqref{Htotbis}. The noise models the interaction with an external environment~\cite{Breuer,jz} or, through the ``imaginary noise trick"~\cite{ref:im} the collapse of the wave function. In this second case, which is the one we are primarily interested in, by setting $\gamma \rightarrow \lambda$ and $N_{\ell} \rightarrow q_{\ell}$ with $\ell=1,2,3$ labeling the three space directions, one recovers the  ``imaginary noise'' version of the QMUPL model~\cite{qmupl1,qmupl2}. If instead one replaces the discrete index $\ell$ with the continuous parameter $\mathbf{x}$, then the discrete sum over $\ell$ becomes an integral over $\mathbf{x}$ and by making the substitutions $N_{\ell}\rightarrow\sum_{j=1}^{N_p}\frac{m_{j}}{m_{0}}g(\mathbf{q}_{j}-\mathbf{x})$ with $N_p$ number of particles of the system and therefore $w_{\ell}\left(t\right)\longrightarrow w\left(\mathbf{x},t\right)$, the ``imaginary noise'' version of the first quantization version of the CSL model is obtained~\cite{csl}.

In view of the perturbative expansion, we write the Hamiltonian $H_{\text{\tiny TOT}}$ in Eq.~\eqref{Htotbis} as the sum of two contributions:
\begin{equation}
H_{\text{\tiny TOT}}=H_0+H_1(t)
\end{equation}
where $H_0=H_{\text{\tiny P}}+H_{\text{\tiny R}}$ is the unperturbed Hamiltonian with known eigenvalues $E_{n}$ and relative eigenvectors $\left|n\right\rangle$, while the remaining term describes the interactions with the electromagnetic field and the noise:
\begin{equation}\label{int}
H_{1}(t)=\sum_{j=1}^{N_p}\left(-\frac{e_j}{m_j}\right) \mathbf{A}(\mathbf{x}_j)\cdot\mathbf{p}_j+\sum_{j=1}^{N_p}\frac{e_{j}^{2}}{2m_j}\mathbf{A}^{2}(\mathbf{x}_j)-\sqrt{\gamma}\hbar\sum_{\ell}N_{\ell}\,w_{\ell}\left(t\right).
\end{equation}
In the following calculation we neglect the term containing $\mathbf{A}^2$ since we only need a result at the lowest perturbative order in $e$ and $\gamma$. Then Eq.~\eqref{int} becomes:
\begin{equation}\label{H_1}
H_{1}(t)=\sum_{j=1}^{N_p}\left(-\frac{e_j}{m_j}\right) \mathbf{A}(\mathbf{x}_j)\cdot\mathbf{p}_j-\sqrt{\gamma}\hbar\sum_{\ell}N_{\ell}\,w_{\ell}\left(t\right).
\end{equation}

\section{The emission rate at the lowest perturbative order}

In order to compute the emission rate, we need the transition probability from the initial state $\left|i;\Omega\right\rangle = \left|i\right\rangle \otimes \left|\Omega\right\rangle$ to the final state $\left|f;\mathbf{k},\lambda\right\rangle = \left|f\right\rangle \otimes \left|\mathbf{k},\lambda\right\rangle$. Here $\left|i\right\rangle$ and $\left|f\right\rangle$ denote, respectively, the initial and final states of the system that are eigenvectors of $H_{\text{\tiny P}}$, while $\left|\Omega\right\rangle$ and $\left|\mathbf{k},\lambda\right\rangle$ denote respectively the vacuum state of the electromagnetic field and the state with one photon with the wave vector $\mathbf{k}$ and polarization $\lambda$. The transition probability is given by:

\begin{equation}\label{eq:6Pfinew}
P_{fi}=\mathbb{E}[\left|\left\langle f;\mathbf{k},\lambda\left|U\left(t,t_{i}\right)\right|i;\Omega\right\rangle \right|^{2}]
\end{equation}
where $U\left(t,t_{i}\right)$ is the time evolution operator from the initial time $t_i$ to the time $t$. Using the standard perturbative approach, we compute the relevant contribution at the lowest order, which is given by the diagrams in Fig. 1.
\begin{center}
{\includegraphics[width=16.5cm, keepaspectratio]{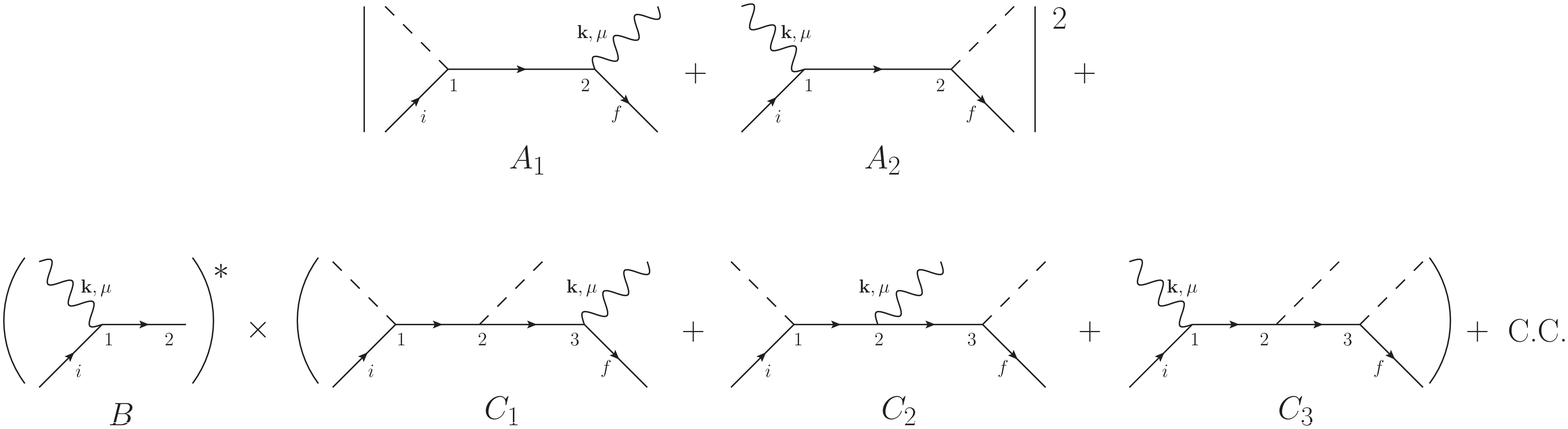}}
\end{center} Fig. 1: Lowest order contributions to the emission rate, represented in terms of Feynman diagrams. Here ``C.C." denotes the complex conjugate of the term in the second line. Solid lines represent the charged fermion, wavy lines the photon, and dashed lines the noise field. In the above diagrams each electromagnetic vertex gives a factor proportional to $e$ while each noise vertex gives a factor proportional to $\sqrt\gamma$.

As shown in appendix A, the corresponding amplitudes are:
\begin{eqnarray}
A_{1} &=& \left(\frac{-i}{\hbar}\right)^{2}\left(-\hbar\sqrt{\gamma}\right)\int_{t_{i}}^{t}dt_{1}\int_{t_{i}}^{t_{1}}dt_{2}\sum_{n}\sum_{\ell}e^{i\left(\Delta_{fn}+\omega_{k}\right)t_{1}}e^{i\Delta_{ni}t_{2}}w_{\ell}\left(t_{2}\right)\left\langle f\left|R_{k}\right|n\right\rangle \left\langle n\left|N_{\ell}\right|i\right\rangle; \label{eq:6A1} \\
\nonumber\\
A_{2} &=&\left(\frac{-i}{\hbar}\right)^{2}\left(-\hbar\sqrt{\gamma}\right)\int_{t_{i}}^{t}dt_{1}\int_{t_{i}}^{t_{1}}dt_{2}\sum_{n}\sum_{\ell}e^{i\Delta_{fn}t_{1}}e^{i\left(\Delta_{ni}+\omega_{k}\right)t_{2}}w_{\ell}\left(t_{1}\right)\left\langle f\left|N_{\ell}\right|n\right\rangle \left\langle n\left|R_{k}\right|i\right\rangle; \label{eq:6A2}\\
\nonumber\\
B &=& \left(\frac{-i}{\hbar}\right)\int_{t_{i}}^{t}dt_{1}e^{i\left(\Delta_{fi}+\omega_{k}\right)t_{1}}\left\langle f\left|R_{k}\right|i\right\rangle; \label{eq:6B} \\
\nonumber\\
C_{1} &=& \left(\frac{-i}{\hbar}\right)^{3}\hbar^{2}\gamma\int_{t_{i}}^{t}dt_{1}\int_{t_{i}}^{t_{1}}dt_{2}\int_{t_{i}}^{t_{2}}dt_{3}\sum_{n,m}\sum_{\ell,\ell'}e^{i\left(\Delta_{fn}+\omega_{k}\right)t_{1}}e^{i\Delta_{nm}t_{2}}e^{i\Delta_{mi}t_{3}}w_{\ell}\left(t_{2}\right)w_{\ell'}\left(t_{3}\right) \times \nonumber \label{eq:6C1}\\
\nonumber\\
&&\;\;\;\;\;\;\;\;\;\;\;\;\;\;\;\;\;\;\;\;\;\;\;\;\;\;\;\;\;\;\;\;\;\;\;\;\;\;\;\;\;\;\;\;\;\;\;\;\;\;\;\;\;\;\;\;\;\;\;\;\;\;\;\;\;\;\;\;\;\;\;\;\;\;\;\;\;\;\; \times \left\langle f\left|R_{k}\right|n\right\rangle \left\langle n\left|N_{\ell}\right|m\right\rangle \left\langle m\left|N_{\ell'}\right|i\right\rangle\\
\nonumber
\end{eqnarray}
\begin{eqnarray}
C_{2} &=& \left(\frac{-i}{\hbar}\right)^{3}\hbar^{2}\gamma\int_{t_{i}}^{t}dt_{1}\int_{t_{i}}^{t_{1}}dt_{2}\int_{t_{i}}^{t_{2}}dt_{3}\sum_{n,m}\sum_{\ell,\ell'}e^{i\Delta_{fn}t_{1}}e^{i\left(\Delta_{nm}+\omega_{k}\right)t_{2}}e^{i\Delta_{mi}t_{3}}w_{\ell}\left(t_{1}\right)w_{\ell'}\left(t_{3}\right) \times \nonumber \label{eq:6C2}\\
\nonumber\\
&&\;\;\;\;\;\;\;\;\;\;\;\;\;\;\;\;\;\;\;\;\;\;\;\;\;\;\;\;\;\;\;\;\;\;\;\;\;\;\;\;\;\;\;\;\;\;\;\;\;\;\;\;\;\;\;\;\;\;\;\;\;\;\;\;\;\;\;\;\;\;\;\;\;\;\;\;\;\;\; \times \left\langle f\left|N_{\ell}\right|n\right\rangle \left\langle n\left|R_{k}\right|m\right\rangle \left\langle m\left|N_{\ell'}\right|i\right\rangle \\
\nonumber\\
C_{3} &=& \left(\frac{-i}{\hbar}\right)^{3}\hbar^{2}\gamma\int_{t_{i}}^{t}dt_{1}\int_{t_{i}}^{t_{1}}dt_{2}\int_{t_{i}}^{t_{2}}dt_{3}\sum_{n,m}\sum_{\ell,\ell'}e^{i\Delta_{fn}t_{1}}e^{i\Delta_{nm}t_{2}}e^{i\left(\Delta_{mi}+\omega_{k}\right)t_{3}}w_{\ell}\left(t_{1}\right)w_{\ell'}\left(t_{2}\right) \times \nonumber \label{eq:6C3}\\
&&\;\;\;\;\;\;\;\;\;\;\;\;\;\;\;\;\;\;\;\;\;\;\;\;\;\;\;\;\;\;\;\;\;\;\;\;\;\;\;\;\;\;\;\;\;\;\;\;\;\;\;\;\;\;\;\;\;\;\;\;\;\;\;\;\;\;\;\;\;\;\;\;\;\;\;\;\;\;\; \times \left\langle f\left|N_{\ell}\right|n\right\rangle \left\langle n\left|N_{\ell'}\right|m\right\rangle \left\langle m\left|R_{k}\right|i\right\rangle.
\end{eqnarray}
Here we have introduced the radiation matrix element:
\begin{equation}\label{eq:6R}
R_{k}:=\;\alpha_{k}\sum_{j=1}^{N_p}\left(-\frac{e_{j}}{m_{j}}\right)e^{-i\mathbf{k}\cdot\mathbf{x}_{j}}\vec{\epsilon}_{\mathbf{k},\lambda}\cdot\mathbf{p}_{j},\qquad\textrm{with}\qquad\alpha_{k}\equiv\sqrt{\frac{\hbar}{2\varepsilon_{0}\omega_{k}\left(2\pi\right)^{3}}}.
\end{equation}
The formula for the transition probability then reads:
\begin{equation}\label{eq:pfiancora}
P_{fi}=\mathbb{E}\left\{ \left|A_{1}+A_{2}\right|^{2}+2\textrm{Re}\left[\left(B^{*}C_{1}+B^{*}C_{2}+B^{*}C_{3}\right)\right]\right\}.
\end{equation}
As anticipated, we are interested in computing the emission rate:
\begin{equation}\label{eq:rategeneral}
\frac{d \Gamma}{dk}=\sum_{\lambda}\int d\Omega_{k}\frac{d}{dt}\sum_{f}P_{fi}\,,
\end{equation}
where, apart doing the sum over the possible final states of the system, we also integrate over the possible directions ($\int d{\Omega}_{k}$) and polarizations ($\sum_{\lambda}$) of the emitted photon. Using Eq.~(\ref{eq:pfiancora}), the emission rate becomes:
\begin{equation}\label{eq:6Rate}
\frac{d \Gamma}{dk}=\sum_{\lambda}\int d\Omega_{k}\frac{d}{dt}\sum_{f}\mathbb{E}\left\{ \left|A_{1}+A_{2}\right|^{2}+2\textrm{Re}\left[\left(B^{*}C_{1}+B^{*}C_{2}+B^{*}C_{3}\right)\right]\right\}. 
\end{equation}
In the next sections we will focus on computing the terms introduced in Eqs.~(\ref{eq:6A1})-(\ref{eq:6C3}). As discussed in the introduction, a direct computation of these terms leads to a wrong result. In the next section we analyze the term $A_1$, pointing out where problems arise and how to avoid them.

\section{The unphysical terms: how to avoid them}
In this section we show that the unphysical contributions to the emission rate arise because of the presence of non resonant terms in the transition amplitude. We will show how to avoid these terms by taking into account the decay of the propagator, due to the electromagnetic self interaction. 

\subsection{The connection with the ``non resonant terms"}
In order to understand the origin of the undesired terms, we study the contribution $A_1$ coming from the diagram ``$A_1$" in Fig. 1. The corresponding transition amplitude for this diagram is given by Eq.~(\ref{eq:6A1}).
In the following analysis it will be convenient to expand the noises $w_{\ell}(t)$ in Fourier components:
\begin{equation} \label{eq:noisefourier}
w_{\ell}(t)=\frac{1}{2\pi}\int_{-\infty}^{+\infty} d\nu\, e^{-i\nu t}w_{\ell}({\nu}),
\end{equation}
so that
\begin{equation} \label{eq:ta_fin62}
A_1 \; = \;\frac{\sqrt{\gamma}}{2\pi\hbar}\underset{n}{\sum}\underset{\ell}{\sum}\left\langle f\left|R_{k}\right|n\right\rangle \left\langle n\left|N_{\ell}\right|i\right\rangle\int_{-\infty}^{+\infty} d\nu\, w_{\ell}({\nu}){\mathrm T}
\end{equation}
with
\begin{eqnarray} \label{eq:TTT6}
{\mathrm T}&:=&\int_{t_i}^{t}dt_{1}
\int_{t_i}^{t_{1}}dt_{2}\,
e^{i(\Delta_{fn}+\omega_{k})t_{1}}
e^{i(\Delta_{ni}-\nu)t_{2}}=\\
\nonumber\\
&=&\int_{t_i}^{t}dt_{2}
\int_{t_2}^{t}dt_{1}\,
e^{i(\Delta_{fn}+\omega_{k})t_{1}}
e^{i(\Delta_{ni}-\nu)t_{2}}.\nonumber
\end{eqnarray}
The emission rate is proportional to the time derivative of the transition probability $P_{fi}=\mathbb{E}\left|A_{1}\right|^{2}$. Using the relation:
\begin{equation}
\mathbb{E}\left[w_{\ell}({\nu})^{*}\,w_{\ell'}({\omega})\right]=2\pi\delta_{{\ell}{\ell'}}\delta\left(\nu-\omega\right)\tilde{f}\left(\nu\right),
\end{equation}
with $\tilde{f}\left(\nu\right)$ defined in Eq.~(\ref{ftilde}), we can write $P_{fi}$ as:
\begin{eqnarray}
P_{fi}&=&\frac{\gamma}{4\pi^{2}\hbar^{2}}\mathbb{E}\left|\underset{n}{\sum}\underset{\ell}{\sum}\left\langle f\left|R_{k}\right|n\right\rangle \left\langle n\left|N_{\ell}\right|i\right\rangle\int_{-\infty}^{+\infty}d\nu\,w_{\ell}({\nu})\mathrm{T}\right|^{2}\label{propropro}\\
\nonumber\\
&=&\frac{\gamma}{2\pi\hbar^{2}}\int_{-\infty}^{+\infty}d\nu\tilde{f}\left(\nu\right)\left|\underset{n}{\sum}\underset{\ell}{\sum}\left\langle f\left|R_{k}\right|n\right\rangle \left\langle n\left|N_{\ell}\right|i\right\rangle\mathrm{T}\right|^{2}.\nonumber
\end{eqnarray}
Let us focus on ${\mathrm T}$, which contains the time dependence of $P_{fi}$, which is the source of the problems. Taking $t_i=0$ we get:
\begin{equation}\label{eq:Texp6}
{\mathrm T}=\frac{-1}{i(\Delta_{fn}+\omega_{k})}\left[\frac{e^{i(\Delta_{fi}+\omega_{k}-\nu)t}-1}{i(\Delta_{fi}+\omega_{k}-\nu)}-e^{i(\Delta_{fn}+\omega_{k})t}\frac{e^{i(\Delta_{ni}-\nu)t}-1}{i(\Delta_{ni}-\nu)}\right]
\end{equation}
When taking the square modulus, in the large time limit the crossed terms oscillate and do not contribute to the emission rate. On the contrary the square modulus of each term in Eq.~(\ref{eq:Texp6}) has the form:
\begin{equation}
\left|\frac{e^{ixt}-1}{ix}\right|^{2}=\frac{\sin^{2}\left(\frac{xt}{2}\right)}{\left(\frac{x}{2}\right)^{2}}\;\;\;\;\underset{t\rightarrow\infty}{\longrightarrow}\;\;\;\;2\pi t\delta\left(x\right).
\end{equation}
The first term in Eq.~(\ref{eq:Texp6}), called resonant term, gives the relevant contribution when the energy is conserved, i.e. when $\nu=\Delta_{fi}+\omega_{k}$. On the contrary the second term in Eq.~(\ref{eq:Texp6}), called non resonant term, becomes relevant when $\nu=\Delta_{ni}$. It is because of this term that, in the case of a free particle, one gets the unphysical contribution to the rate, proportional to ${\tilde f} (0)$ (see Eq.~(\ref{eq:rate non white})).
Notice that the presence of non resonant terms is not related to the fact that our interaction is a noise: they appear also with a generic potential~\cite{Sakurai}. In the next subsection we introduce the decay of the propagator and we show why, by taking this effect into account, the non resonant terms can be neglected.

\subsection{Decay of the propagator}

In~\cite{bd,dirk} it was shown that the higher order contributions of the electromagnetic interaction play a fundamental role in avoiding the presence of the unphysical terms. This suggests that their effect should be introduced also in the perturbative calculations. In particular, it was shown that the role of higher order contributions of the electromagnetic interaction is to exponentially damp the unphysical terms. This exponential damping resembles the exponential decay of the eigenstates of the unperturbed Hamiltonian $H_0$ due to the electromagnetic interaction. In fact, it is well known that, due to the electromagnetic interaction, the eigenstates of the free Hamiltonian $H_0$ are not stable and can decay~\cite{Sakurai,GreinerSPECIAL}. More precisely, the higher order contributions of the electromagnetic interaction add a complex shift $\Delta E=\Delta E_r+i\Delta E_i$ to the eigenenergies. The real part $\Delta E_r$ is a shift of the energy levels (the Lamb shift) while the imaginary part $\Delta E_i$ described the decay rate (or natural broadening) of the state. As a check that this decay is related to the exponential damping factors found in~\cite{bd,dirk}, in appendix B we compute $\Delta E_i$ for an harmonic oscillator. The broadening is shown to be proportional to the decay rate $\Lambda=\frac{\omega_0^2 \beta}{2m}$ found in~\cite{bd,dirk}, responsible for suppressing the terms proportional to ${\tilde f}(0)$. This strongly suggests that the decay of the eigenstates due to electromagnetic interactions plays a fundamental role in avoiding the presence of the unphysical terms\footnote{This is also discussed in~\cite{Sakurai2}, for Compton scattering.}. 

Therefore, we compute again ${\mathrm T}$ of Eq.~\eqref{eq:TTT6} taking into account the possibility that the propagator decays. This is equivalent to replacing, in the integral, $e^{\frac{i}{\hbar}E_n(t_1-t_2)}$ with $e^{\frac{i}{\hbar}(E_n+i \hbar\Gamma_n)(t_1-t_2)}$. In such a case:

\begin{eqnarray}
{\mathrm T} &=& \int_{0}^{t}dt_{1}\int_{0}^{t_{1}}dt_{2}\, e^{[i(\Delta_{fn}+\omega_{k})]t_{1}}e^{[i(\Delta_{ni}-\nu)]t_{2}}e^{-\Gamma_{n}(t_{1}-t_{2})}= \\
& =& \frac{-1}{[i(\Delta_{fn}+\omega_{k})-\Gamma_{n}]}\left\{ \frac{e^{i(\Delta_{fi}+\omega_{k}-\nu)t}-1}{i(\Delta_{fi}+\omega_{k}-\nu)}-e^{[i(\Delta_{fn}+\omega_{k})-\Gamma_{n}]t}\frac{e^{[i(\Delta_{ni}-\nu)+\Gamma_{n}]t}-1}{[i(\Delta_{ni}-\nu)+\Gamma_{n}]}\right\} .\nonumber
\end{eqnarray}

The first term is the same as in Eq.~\eqref{eq:Texp6}. Therefore we still have a contribution in the large time limit, when $\nu=\Delta_{fi}+\omega_{k}$. However, because of the damping $e^{-\Gamma_n t}$, the second term does not contribute anymore to the emission rate for large times. This shows that, taking into account the decay of propagator, one can avoid the presence of the non resonant term and therefore the unphysical factor ${\tilde f}(0)$. The great advantage of this method is that it is quite general and does not depend on the form of the matrix elements $\left\langle f\left|R_{k}\right|n\right\rangle$ and $\left\langle n\left|N_{\ell}\right|i\right\rangle$.

In the rest of the article we apply this method to compute all contributions in Eqs.~(\ref{eq:6A1})-(\ref{eq:6C3}). We will show that, with a proper use of the decay of propagator, the unphysical term is not present anymore in the final formula.

\section{Contributions of the amplitudes $A_1$ and $A_2$ to the emission rate} 
The contribution of the amplitudes $A_1$ and $A_2$ to the emission rate (Eq.~(\ref{eq:6Rate})) can be written as the sum of three terms 
\begin{equation}
\sum_{\lambda}\int d\Omega_{k}\frac{d}{dt}\sum_{f}\mathbb{E}\left|A_{1}+A_{2}\right|^{2}=\frac{\gamma}{\hbar^{2}}\sum_{\lambda}\int d\Omega_{k}\left[R_{11}+2\textrm{Re}\left(R_{12}\right)+R_{22}\right],
\end{equation}
where we have introduced:
\begin{eqnarray}
R_{11} &= &\frac{d}{dt}\sum_{f}\mathbb{E}\left|\sum_{n}\sum_{\ell}\int_{t_{i}}^{t}dt_{1}\int_{t_{i}}^{t_{1}}dt_{2}e^{i\left(\triangle_{fn}+\omega_{k}\right)t_{1}}e^{i\triangle_{ni}t_{2}}w_{\ell}\left(t_{2}\right)\left\langle f\left|R_{k}\right|n\right\rangle \left\langle n\left|N_{\ell}\right|i\right\rangle \right|^{2};\;\;\;\;\;\;\;\;\;\;\;\\
\nonumber\\
R_{12} &= &\frac{d}{dt}\sum_{f}\mathbb{E}\left(\sum_{n}\sum_{\ell}\int_{t_{i}}^{t}dt_{1}\int_{t_{i}}^{t_{1}}dt_{2}e^{i\left(\triangle_{fn}+\omega_{k}\right)t_{1}}e^{i\triangle_{ni}t_{2}}w_{\ell}\left(t_{2}\right)\left\langle f\left|R_{k}\right|n\right\rangle \left\langle n\left|N_{\ell}\right|i\right\rangle \right) \times \nonumber\\
&&\times\left(\sum_{n'}\sum_{\ell'}\int_{t_{i}}^{t}dt_{1}\int_{t_{i}}^{t_{1}}dt_{2}e^{i\triangle_{fn'}t_{1}}e^{i\left(\triangle_{n'i}+\omega_{k}\right)t_{2}}w_{\ell'}\left(t_{1}\right)\left\langle f\left|N_{\ell'}\right|n'\right\rangle \left\langle n'\left|R_{k}\right|i\right\rangle \right)^{*};\\
\nonumber\\
R_{22}&=& \frac{d}{dt}\sum_{f}\mathbb{E}\left|\sum_{n}\sum_{\ell}\int_{t_{i}}^{t}dt_{1}\int_{t_{i}}^{t_{1}}dt_{2}e^{i\triangle_{fn}t_{1}}e^{i\left(\triangle_{ni}+\omega_{k}\right)t_{2}}w_{\ell}\left(t_{1}\right)\left\langle f\left|N_{\ell}\right|n\right\rangle \left\langle n\left|R_{k}\right|i\right\rangle \right|^{2}.
\end{eqnarray}
$R_{11}$ ($R_{22}$) is the transition probability corresponding to the amplitude represented by diagram $A_1$ ($A_2$) in Fig.1. The interference effects between these two transition amplitudes amplitude are contained in the term $R_{12}$. We compute the three terms separately.

\subsection{Computation of $R_{11}$}
Since we will focus on the time dependent part, it is convenient to write $R_{11}$ in the following way:
\begin{equation}\label{R11}
R_{11}=\sum_{f}\sum_{n,m}\sum_{\ell,\ell'}\left\langle f\left|R_{k}\right|n\right\rangle \left\langle n\left|N_{\ell}\right|i\right\rangle \left\langle f\left|R_{k}\right|m\right\rangle ^{*}\left\langle m\left|N_{\ell'}\right|i\right\rangle ^{*}\frac{d}{dt}T_{1},
\end{equation}
where:
\begin{equation}\label{eq:6T1}
T_{1}:=\int_{t_{i}}^{t}dt_{1}\int_{t_{i}}^{t_{1}}dt_{2}\int_{t_{i}}^{t}dt_{3}\int_{t_{i}}^{t_{3}}dt_{4}e^{i\left(\triangle_{fn}+\omega_{k}\right)t_{1}}e^{i\triangle_{ni}t_{2}}e^{-i\left(\triangle_{fm}+\omega_{k}\right)t_{3}}e^{-i\triangle_{mi}t_{4}}\mathbb{E}\left[w_{\ell}\left(t_{2}\right)w_{\ell'}\left(t_{4}\right)\right].
\end{equation}
Until now we never introduced the decay of the propagator. If one computes $T_1$ as defined in Eq.~(\ref{eq:6T1}) then the unphysical term is present. Taking into account the decay of the propagator, i.e., the fact that the intermediate states $\left|m\right\rangle$ and $\left|n\right\rangle$ decay respectively with rates $\Gamma_m$ and $\Gamma_n$, amounts to introducing the exponentials $e^{-\Gamma_{m}(t_3-t_4)}$ and $e^{-\Gamma_{n}(t_1-t_2)}$ in Eq.~(\ref{eq:6T1}). Then, setting for simplicity $t_{i}=0$ and using $\mathbb{E}\left[w_{\ell}\left(t_{2}\right)w_{\ell'}\left(t_{4}\right)\right]=\delta_{\ell,\ell'}f(t_2-t_4)$, Eq.~(\ref{eq:6T1}) becomes:
\[
T_{1}=\delta_{\ell,\ell'}\int_{0}^{t}dt_{1}\int_{0}^{t_{1}}dt_{2}\int_{0}^{t}dt_{3}\int_{0}^{t_{1}}dt_{4}e^{i\left(\triangle_{fn}+\omega_{k}\right)t_{1}}e^{i\triangle_{ni}t_{2}}e^{-i\left(\triangle_{fm}+\omega_{k}\right)t_{3}}e^{-i\triangle_{mi}t_{4}}f\left(t_{2}-t_{4}\right)\times
\]
\[
\times e^{-\Gamma_{n}\left(t_{1}-t_{2}\right)}e^{-\Gamma_{m}\left(t_{3}-t_{4}\right)}= 
\]
\[
=\delta_{\ell,\ell'}\int_{0}^{t}dt_{2}\int_{0}^{t}dt_{4}e^{\left(i\triangle_{ni}+\Gamma_{n}\right)t_{2}}e^{\left(-i\triangle_{mi}+\Gamma_{m}\right)t_{4}}f\left(t_{2}-t_{4}\right)\left(\int_{t_{2}}^{t}dt_{1}e^{\left[i\left(\triangle_{fn}+\omega_{k}\right)-\Gamma_{n}\right]t_{1}}\right)\times
\]
\[
\;\;\;\;\;\;\;\;\;\;\;\;\times \left(\int_{t_{4}}^{t}dt_{3}e^{\left[-i\left(\triangle_{fm}+\omega_{k}\right)-\Gamma_{m}\right]t_{3}}\right)=
\]
\begin{eqnarray}\label{eq:6T1new}
&=&\frac{\delta_{\ell,\ell'}}{\left[i\left(\triangle_{fn}+\omega_{k}\right)-\Gamma_{n}\right]\left[-i\left(\triangle_{fm}+\omega_{k}\right)-\Gamma_{m}\right]}\times \nonumber\\
\nonumber\\
&\times& \left\{e^{\left[i\left(\triangle_{mn}\right)-\left(\Gamma_{n}+\Gamma_{m}\right)\right]t}\int_{0}^{t}dt_{2}\int_{0}^{t}dt_{4}e^{\left(i\triangle_{ni}+\Gamma_{n}\right)t_{2}}e^{\left(-i\triangle_{mi}+\Gamma_{m}\right)t_{4}}f\left(t_{2}-t_{4}\right)\right.\nonumber\\
\nonumber\\
& -&e^{\left[i\left(\triangle_{fn}+\omega_{k}\right)-\Gamma_{n}\right]t}\int_{0}^{t}dt_{2}\int_{0}^{t}dt_{4}e^{\left(i\triangle_{ni}+\Gamma_{n}\right)t_{2}}e^{-i\left(\triangle_{fi}+\omega_{k}\right)t_{4}}f\left(t_{2}-t_{4}\right) \nonumber \\
\nonumber \\
&-&e^{\left[-i\left(\triangle_{fm}+\omega_{k}\right)-\Gamma_{m}\right]t}\int_{0}^{t}dt_{2}\int_{0}^{t}dt_{4}e^{i\left(\triangle_{fi}+\omega_{k}\right)t_{2}}e^{\left(-i\triangle_{mi}+\Gamma_{m}\right)t_{4}}f\left(t_{2}-t_{4}\right) \nonumber \\
\nonumber \\
&+&\left.\int_{0}^{t}dt_{2}\int_{0}^{t}dt_{4}e^{i\left(\triangle_{fi}+\omega_{k}\right)\left(t_{2}-t_{4}\right)}f\left(t_{2}-t_{4}\right)\right\}. 
\end{eqnarray}   
In the large time limit $t\rightarrow\infty$ only the last term survives. Notice that without the introduction of the decay of the propagator, also the first term in Eq.~(\ref{eq:6T1new}) would have not been negligible, giving rise to a term proportional to $\tilde{f}(0)$. 
Using the relation:
\begin{equation}\label{eq:6timeint}
\frac{d}{dt}e^{-\left(a+b\right)t}\left(\int_{0}^{t}dt_{1}\int_{0}^{t}dt_{2}e^{at_{1}}e^{bt_{2}}f\left(t_{1}-t_{2}\right)\right)=e^{-\left(a+b\right)t}\left(\int_{0}^{t}dxe^{ax}f\left(x\right)+\int_{0}^{t}dxe^{bx}f\left(x\right)\right).
\end{equation}
and the fact that in the fourth line of Eq.~(\ref{eq:6T1new}) we have $a=-b=\triangle_{fi}+\omega_{k}$, we get, in the large time limit,
\begin{equation}
\frac{d}{dt}T_{1}=\frac{\delta_{\ell,\ell'}}{\left[i\left(\triangle_{fn}+\omega_{k}\right)-\Gamma_{n}\right]\left[-i\left(\triangle_{fm}+\omega_{k}\right)-\Gamma_{m}\right]}\tilde{f}\left(\triangle_{fi}+\omega_{k}\right),
\end{equation}
with $\tilde{f}(k)$ defined in Eq.~(\ref{ftilde}). Coming back to Eq.~(\ref{R11}), we have:
\begin{equation}\label{eq:6R1def}
R_{11}=\sum_{f}\sum_{n,m}\sum_{\ell}\frac{\left\langle f\left|R_{k}\right|n\right\rangle \left\langle n\left|N_{\ell}\right|i\right\rangle \left\langle f\left|R_{k}\right|m\right\rangle ^{*}\left\langle m\left|N_{\ell}\right|i\right\rangle ^{*}}{\left[i\left(\triangle_{fn}+\omega_{k}\right)-\Gamma_{n}\right]\left[-i\left(\triangle_{fm}+\omega_{k}\right)-\Gamma_{m}\right]}\tilde{f}\left(\triangle_{fi}+\omega_{k}\right).
\end{equation}

\subsection{Computation of $R_{12}$}

Similarly to the computation of $R_{11}$, we start by splitting the time dependent part of $R_{12}$ from the rest:
\begin{equation}\label{R12}
R_{2}=\sum_{f}\sum_{n,m}\sum_{\ell}\left\langle f\left|R_{k}\right|n\right\rangle \left\langle n\left|N_{\ell}\right|i\right\rangle \left\langle f\left|N_{\ell}\right|m\right\rangle ^{*}\left\langle m\left|R_{k}\right|i\right\rangle ^{*}\frac{d}{dt}T_2
\end{equation}
where
\begin{equation}
T_{2}:=\int_{0}^{t}dt_{1}\int_{0}^{t_{1}}dt_{2}\int_{0}^{t}dt_{3}\int_{0}^{t_{3}}dt_{4}e^{i\left(\triangle_{fn}+\omega_{k}\right)t_{1}}e^{i\triangle_{ni}t_{2}}e^{-i\triangle_{fm}t_{3}}e^{-i\left(\triangle_{mi}+\omega_{k}\right)t_{4}}f\left(t_{2}-t_{3}\right).
\end{equation}
As before, we introduce the decay of the propagator by including the exponentials $e^{-\Gamma_{n}\left(t_{1}-t_{2}\right)}$ and $e^{-\Gamma_{m}\left(t_{3}-t_{4}\right)}$ in the equation above (we also set $t_{i}=0$):
\[
T_{2}=\int_{0}^{t}dt_{1}\int_{0}^{t_{1}}dt_{2}\int_{0}^{t}dt_{3}\int_{0}^{t_{3}}dt_{4}e^{i\left(\triangle_{fn}+\omega_{k}\right)t_{1}}e^{i\triangle_{ni}t_{2}}e^{-i\triangle_{fm}t_{3}}e^{-i\left(\triangle_{mi}+\omega_{k}\right)t_{4}}f\left(t_{2}-t_{3}\right)\times
\]
\[
\times e^{-\Gamma_{n}\left(t_{1}-t_{2}\right)}e^{-\Gamma_{m}\left(t_{3}-t_{4}\right)}=
\]
\[
=\int_{0}^{t}dt_{2}\int_{0}^{t}dt_{3}\left(\int_{t_{2}}^{t}dt_{1}e^{\left[i\left(\triangle_{fn}+\omega_{k}\right)-\Gamma_{n}\right]t_{1}}\right)e^{\left(i\triangle_{ni}+\Gamma_{n}\right)t_{2}}e^{\left(-i\triangle_{fm}-\Gamma_{m}\right)t_{3}}\times\;\;\;\;\;\;\;\;\;\;\;\;\;\;\;\;
\]
\[
\;\;\;\;\;\;\;\;\;\;\;\;\;\;\;\;\;\;\;\;\;\;\;\;\;\;\;\;\times\left(\int_{0}^{t_{3}}dt_{4}e^{\left[-i\left(\triangle_{mi}+\omega_{k}\right)+\Gamma_{m}\right]t_{4}}\right)f\left(t_{2}-t_{3}\right)=
\]
\begin{eqnarray}\label{eq:6T2new}
&=& \frac{1}{\left[i\left(\triangle_{fn}+\omega_{k}\right)-\Gamma_{n}\right]\left[-i\left(\triangle_{mi}+\omega_{k}\right)+\Gamma_{m}\right]}\nonumber\\
\nonumber\\
&\times&\left\{ e^{\left[i\left(\triangle_{fn}+\omega_{k}\right)-\Gamma_{n}\right]t}\int_{0}^{t}dt_{2}\int_{0}^{t}dt_{3}e^{\left(i\triangle_{ni}+\Gamma_{n}\right)t_{2}}e^{-i\left(\triangle_{fi}+\omega_{k}\right)t_{3}}f\left(t_{2}-t_{3}\right)\right.\nonumber\\
\nonumber\\
& 
-&e^{\left[i\left(\triangle_{fn}+\omega_{k}\right)-\Gamma_{n}\right]t}\int_{0}^{t}dt_{2}\int_{0}^{t}dt_{3}e^{\left(i\triangle_{ni}+\Gamma_{n}\right)t_{2}}e^{\left(-i\triangle_{fm}-\Gamma_{m}\right)t_{3}}f\left(t_{2}-t_{3}\right) \nonumber\\
\nonumber\\
&-&\int_{0}^{t}dt_{2}\int_{0}^{t}dt_{3}e^{i\left(\triangle_{fi}+\omega_{k}\right)\left(t_{2}-t_{3}\right)}f\left(t_{2}-t_{3}\right)\nonumber\\
\nonumber\\
&+&\left.\int_{0}^{t}dt_{2}\int_{0}^{t}dt_{3}e^{i\left(\triangle_{fi}+\omega_{k}\right)t_{2}}e^{\left(-i\triangle_{fm}-\Gamma_{m}\right)t_{3}}f\left(t_{2}-t_{3}\right)\right\} 
\end{eqnarray}
The first two terms of Eq.~(\ref{eq:6T2new}) have the same structure as in Eq.~(\ref{eq:6timeint}) and, in the large time limit, they go to zero. The third and the fourth term of Eq.~(\ref{eq:6T2new}) have the structure: 
\begin{eqnarray}\label{eq:6Iabt}
I(a,b,t):&=&\int_{0}^{t}dt_{1}\int_{0}^{t}dt_{2}e^{at_{1}}e^{bt_{2}}f\left(t_{1}-t_{2}\right) \\
\nonumber\\
&=&\frac{1}{a+b}\left(e^{\left(a+b\right)t}\int_{0}^{t}dxe^{-bx}f\left(x\right)-\int_{0}^{t}dxe^{ax}f\left(x\right)+e^{\left(a+b\right)t}\int_{0}^{t}dxe^{-ax}f\left(x\right)-\int_{0}^{t}dxe^{bx}f\left(x\right)\right). \nonumber
\end{eqnarray}
The emission rate is proportional to the time derivative of $I(a,b,t)$. In the large time limit the only terms which survive are those with $a=-b$. In fact, in such a case $I(a,b,t)$ increases linearly with time:
\begin{equation}
I(a,-a,t)=\int_{0}^{t}dxe^{ax}\left(t-x\right)f\left(x\right)+\int_{0}^{t}dx\left(t-x\right)e^{-ax}f\left(x\right)\quad\underset{t\rightarrow\infty}{\sim}\quad t\int_{-t}^{t}dxe^{ax}f\left(x\right),
\end{equation} 
while for $a\neq-b$ it oscillates or goes to zero. Then only the third term in Eq.~(\ref{eq:6T2new}) survives so that:
\begin{equation}
\frac{d}{dt}T_{2}\quad\underset{t\rightarrow\infty}{\longrightarrow}\quad\frac{\tilde{f}\left(\triangle_{fi}+\omega_{k}\right)}{\left[i\left(\triangle_{fn}+\omega_{k}\right)-\Gamma_{n}\right]\left[-i\left(\triangle_{mi}+\omega_{k}\right)+\Gamma_{m}\right]}.
\end{equation}
Coming back to Eq.~(\ref{R12}) we have:
\begin{equation}\label{eq:6R2def}
R_{12}=\sum_{f}\sum_{n,m}\sum_{\ell}\frac{\left\langle f\left|R_{k}\right|n\right\rangle \left\langle n\left|N_{\ell}\right|i\right\rangle \left\langle f\left|N_{\ell}\right|m\right\rangle ^{*}\left\langle m\left|R_{k}\right|i\right\rangle ^{*}}{\left[i\left(\triangle_{fn}+\omega_{k}\right)-\Gamma_{n}\right]\left[-i\left(\triangle_{mi}+\omega_{k}\right)+\Gamma_{m}\right]}\tilde{f}\left(\triangle_{fi}+\omega_{k}\right).
\end{equation}

\subsection{Computation of $R_{22}$}
As in the previous cases, we write $R_{22}$ as:
\begin{equation}
R_{22}=\sum_{f}\sum_{m,n}\sum_{\ell}\left\langle f\left|N_{\ell}\right|n\right\rangle \left\langle n\left|R_{k}\right|i\right\rangle \left\langle f\left|N_{\ell}\right|m\right\rangle ^{*}\left\langle m\left|R_{k}\right|i\right\rangle ^{*}\frac{d}{dt}T_{3}\,,
\end{equation}
where
\begin{equation}
T_{3}:=\int_{t_{i}}^{t}dt_{1}\int_{t_{i}}^{t_{1}}dt_{2}\int_{t_{i}}^{t}dt_{3}\int_{t_{i}}^{t_{3}}dt_{4}e^{i\triangle_{fn}t_{1}}e^{i\left(\triangle_{ni}+\omega_{k}\right)t_{2}}f\left(t_{1}-t_{3}\right)e^{-i\triangle_{fm}t_{3}}e^{-i\left(\triangle_{mi}+\omega_{k}\right)t_{4}}.
\end{equation}
By adding the decay of the propagator and setting $t_{i}=0$ we have:
\[
T_{3}=\int_{0}^{t}dt_{1}\int_{0}^{t_{1}}dt_{2}\int_{0}^{t}dt_{3}\int_{0}^{t_{3}}dt_{4}e^{i\triangle_{fn}t_{1}}e^{i\left(\triangle_{ni}+\omega_{k}\right)t_{2}}f\left(t_{1}-t_{3}\right)e^{-i\triangle_{fm}t_{3}}e^{-i\left(\triangle_{mi}+\omega_{k}\right)t_{4}}\times
\]
\[
\times e^{-\Gamma_{n}\left(t_{1}-t_{2}\right)}e^{-\Gamma_{m}\left(t_{3}-t_{4}\right)}=
\]
\[
=\int_{0}^{t}dt_{1}\int_{0}^{t}dt_{3}e^{\left(i\triangle_{fn}-\Gamma_{n}\right)t_{1}}\left(\int_{0}^{t_{1}}dt_{2}e^{\left[i\left(\triangle_{ni}+\omega_{k}\right)+\Gamma_{n}\right]t_{2}}\right)f\left(t_{1}-t_{3}\right)e^{\left(-i\triangle_{fm}-\Gamma_{m}\right)t_{3}}\times
\]
\[
\;\;\;\;\;\;\;\;\;\;\;\times \left(\int_{0}^{t_{3}}dt_{4}e^{\left[-i\left(\triangle_{mi}+\omega_{k}\right)+\Gamma_{m}\right]t_{4}}\right)=
\]
\begin{eqnarray}\label{eq:6T3new}
=&&\frac{1}{\left[i\left(\triangle_{ni}+\omega_{k}\right)+\Gamma_{n}\right]\left[-i\left(\triangle_{mi}+\omega_{k}\right)+\Gamma_{m}\right]}\left\{ \int_{0}^{t}dt_{1}\int_{0}^{t}dt_{3}e^{i\left(\triangle_{fi}+\omega_{k}\right)\left(t_{1}-t_{3}\right)}f\left(t_{1}-t_{3}\right)\right.\nonumber\\
\nonumber\\
-&&\int_{0}^{t}dt_{1}\int_{0}^{t}dt_{3}e^{i\left(\triangle_{fi}+\omega_{k}\right)t_{1}}e^{\left(-i\triangle_{fm}-\Gamma_{m}\right)t_{3}}f\left(t_{1}-t_{3}\right) \nonumber\\
\nonumber\\
-&&\int_{0}^{t}dt_{1}\int_{0}^{t}dt_{3}e^{\left(i\triangle_{fn}-\Gamma_{n}\right)t_{1}}\left(e^{-i\left(\triangle_{fi}+\omega_{k}\right)t_{3}}\right)f\left(t_{1}-t_{3}\right)\nonumber\\
\nonumber\\
+&&\left.\int_{0}^{t}dt_{1}\int_{0}^{t}dt_{3}e^{\left(i\triangle_{fn}-\Gamma_{n}\right)t_{1}}e^{\left(-i\triangle_{fm}-\Gamma_{m}\right)t_{3}}f\left(t_{1}-t_{3}\right)\right\}.
\end{eqnarray} 
All four terms in Eq.~\eqref{eq:6T3new} contain integrals with the same structure of $I(a,b,t)$ given in Eq.~\eqref{eq:6Iabt}. As already discussed, the only relevant contribution in the large time limit is that with $a=-b$. Then only the first term in Eq.~\eqref{eq:6T3new} contributes, which implies:
\begin{equation} 
\frac{d}{dt}T_{3}\quad\underset{t\rightarrow\infty}{\longrightarrow}\quad\frac{\tilde{f}\left(\triangle_{fi}+\omega_{k}\right)}{\left[i\left(\triangle_{ni}+\omega_{k}\right)+\Gamma_{n}\right]\left[-i\left(\triangle_{mi}+\omega_{k}\right)+\Gamma_{m}\right]}.
\end{equation}
Thus we have:
\begin{equation}\label{eq:6R3def} 
R_{22}=\sum_{f}\sum_{m,n}\sum_{\ell}\frac{\left\langle f\left|N_{\ell}\right|n\right\rangle \left\langle n\left|R_{k}\right|i\right\rangle \left\langle f\left|N_{\ell}\right|m\right\rangle ^{*}\left\langle m\left|R_{k}\right|i\right\rangle ^{*}}{\left[i\left(\triangle_{ni}+\omega_{k}\right)+\Gamma_{n}\right]\left[-i\left(\triangle_{mi}+\omega_{k}\right)+\Gamma_{m}\right]}\tilde{f}\left(\triangle_{fi}+\omega_{k}\right).
\end{equation}

\section{Contribution due to the mixed terms}
In this section we study the contribution to the rate due to the mixed terms $B^*C_1$, $B^*C_2$ and $B^*C_3$ (see Eq.~\eqref{eq:6Rate}).
\subsection{Computation of $B^*C_1$}
We start from the contribution from the terms $B$ and $C_1$ given in Eqs.~\eqref{eq:6B} and \eqref{eq:6C1}. The $B$ term is simply:
\begin{equation}
B=\left(\frac{-i}{\hbar}\right)\left\langle f\left|R_{k}\right|i\right\rangle T_B
\end{equation}
where, setting $t_i=0$, 
\begin{equation}\label{eq:6TB}
T_B=\int_{0}^{t}dt_{1}e^{i\left(\triangle_{fi}+\omega_{k}\right)t_{1}}=\frac{e^{i\left(\triangle_{fi}+\omega_{k}\right)t}-1}{i\left(\triangle_{fi}+\omega_{k}\right)}.
\end{equation}
The term $C_1$ can be written as:
\begin{equation}
C_{1}=\frac{i\gamma}{\hbar}\sum_{n,m}\sum_{\ell,\ell'}\left\langle f\left|R_{k}\right|n\right\rangle \left\langle n\left|N_{\ell}\right|m\right\rangle \left\langle m\left|N_{\ell'}\right|i\right\rangle T_{C_{1}}. 
\end{equation}
After taking the average over the noise, $T_{C_{1}}$ becomes:
\begin{eqnarray}
T_{C_{1}}&=&\delta_{\ell,\ell'}\int_{0}^{t}\!\!\!dt_{1}\int_{0}^{t_{1}}\!\!\!dt_{2}\int_{0}^{t_{2}}\!\!\!dt_{3}\,e^{i\left(\triangle_{fn}+\omega_{k}\right)t_{1}}e^{i\triangle_{nm}t_{2}}e^{i\triangle_{mi}t_{3}}f\left(t_{2}-t_{3}\right)e^{-\Gamma_{n}\left(t_{1}-t_{2}\right)}e^{-\Gamma_{m}\left(t_{2}-t_{3}\right)}\\
\nonumber\\
&=&\int_{0}^{t}\!\!\!dt_{1}\,e^{i\left(\triangle_{fn}+\omega_{k}\right)t_{1}}e^{-\Gamma_{n}t_{1}}\int_{0}^{t_{1}}\!\!\!dt_{2}\int_{0}^{t_{1}}\!\!\!dt_{3}\,\theta\left(t_{2}-t_{3}\right)e^{i\triangle_{nm}t_{2}}e^{\Gamma_{n}t_{2}}e^{i\triangle_{mi}t_{3}}f\left(t_{2}-t_{3}\right)e^{-\Gamma_{m}\left(t_{2}-t_{3}\right)}.\nonumber
\end{eqnarray}
The integral can be computed in a way similar to the previous cases and the final result is:
\begin{eqnarray}
T_{C_{1}}&=&\frac{\delta_{\ell,\ell'}}{\left(i\triangle_{ni}+\Gamma_{n}\right)i\left(\triangle_{fi}+\omega_{k}\right)}\times \nonumber\\
\nonumber\\
&\times&\left(e^{i\left(\triangle_{fi}+\omega_{k}\right)t}\int_{0}^{t}dxe^{\left(i\triangle_{im}-\Gamma_{m}\right)x}f\left(x\right)-\int_{0}^{t}dxf\left(x\right)e^{\left[i\left(\triangle_{fm}+\omega_{k}\right)-\Gamma_{m}\right]x}\right)-\nonumber\\
\nonumber\\
&-&\frac{\delta_{\ell,\ell'}}{\left(i\triangle_{ni}+\Gamma_{n}\right)\left[i\left(\triangle_{fn}+\omega_{k}\right)-\Gamma_{n}\right]} \times \nonumber\\
\nonumber\\
&\times&\left(e^{\left[i\left(\triangle_{fn}+\omega_{k}\right)-\Gamma_{n}\right]t}\int_{0}^{t}dxe^{\left[i\triangle_{nm}+\left(\Gamma_{n}-\Gamma_{m}\right)\right]x}f\left(x\right)-\int_{0}^{t}dxf\left(x\right)e^{\left[i\left(\triangle_{fm}+\omega_{k}\right)-\Gamma_{m}\right]x}\right)
\end{eqnarray}
Since in the formula for the rate (Eq.~\eqref{eq:6Rate}) we need to compute $\mathbb{E}\left\{ 2\textrm{Re}\left(B^{*}C_{1}\right)\right\} $, we focus on: 
\begin{equation}
T_{B{C_{1}}}=T_{B}\cdot T_{C_{1}}=\frac{T_{C_{1}}-T_{C_{1}}e^{-i\left(\triangle_{fi}+\omega_{k}\right)t}}{i\left(\triangle_{fi}+\omega_{k}\right)}
\end{equation}
and in particular on its time derivative:
\begin{equation}
\frac{d}{dt}T_{BC_{1}}=\frac{\frac{d}{dt}T_{C_{1}}-\frac{d}{dt}\left[T_{C_{1}}e^{-i\left(\triangle_{fi}+\omega_{k}\right)t}\right]}{i\left(\triangle_{fi}+\omega_{k}\right)}.
\end{equation}
It is straightforward to see that, in the large time limit:
\begin{eqnarray}
\frac{dT_{C_{1}}}{dt}&\quad\underset{t\rightarrow\infty}{\sim}\quad & \delta_{\ell,\ell'}\frac{e^{i\left(\triangle_{fi}+\omega_{k}\right)t}}{\left(i\triangle_{ni}+\Gamma_{n}\right)}\int_{0}^{t}dxe^{\left(i\triangle_{im}-\Gamma_{m}\right)x}f\left(x\right),\nonumber\\
\nonumber\\
\frac{d}{dt}\left[T_{C_{1}}e^{-i\left(\triangle_{fi}+\omega_{k}\right)t}\right] &\quad\underset{t\rightarrow\infty}{\sim}\quad & -\delta_{\ell,\ell'}\frac{e^{-i\left(\triangle_{fi}+\omega_{k}\right)t}}{\left[i\left(\triangle_{fn}+\omega_{k}\right)-\Gamma_{n}\right]}\int_{0}^{t}dxf\left(x\right)e^{\left[i\left(\triangle_{fm}+\omega_{k}\right)-\Gamma_{m}\right]x} \nonumber
\end{eqnarray}
Then in the large time limit:
\begin{eqnarray}\label{eq:6bobo}
\frac{d}{dt}T_{BC_{1}}&=&\delta_{\ell,\ell'}\frac{1}{i\left(\triangle_{fi}+\omega_{k}\right)}\times \left\{\frac{e^{i\left(\triangle_{fi}+\omega_{k}\right)t}}{\left(i\triangle_{ni}+\Gamma_{n}\right)}\int_{0}^{t}dxe^{\left(i\triangle_{im}-\Gamma_{m}\right)x}f\left(x\right)+\right.\nonumber\\
\nonumber\\
&+&\left.\frac{e^{-i\left(\triangle_{fi}+\omega_{k}\right)t}}{\left[i\left(\triangle_{fn}+\omega_{k}\right)-\Gamma_{n}\right]}\int_{0}^{t}dxf\left(x\right)e^{\left[i\left(\triangle_{fm}+\omega_{k}\right)-\Gamma_{m}\right]x}\right\}
\end{eqnarray}
The important point is that Eq.~\eqref{eq:6bobo} contains oscillating terms, which average to zero for $t\rightarrow\infty$. The only exception is when $\triangle_{fi}+\omega_{k}=0$.
However, so long as we study systems whose initial state is the ground state, the condition $\triangle_{fi}+\omega_{k}=0$ is never fulfilled. 

When the initial state is not the ground state, the effect of the spontaneous emission due only to the vacuum fluctuations of the electromagnetic field is expected to be bigger than the one given in Eq.~\eqref{eq:6bobo}, which is due to the presence of the noise. This can be qualitatively understood by considering the fact that vacuum fluctuations are of order $e^2$, while the effect given by Eq.~\eqref{eq:6bobo} is of order $e^2\gamma$. Therefore, before having any emission due to the noise, the system is already decayed in its ground state because of the vacuum fluctuations of the electromagnetic field. Since we are interested only in the radiation emission induced by the noise, this contribution can be neglected. 

The same conclusions are true also for the contributions $B^*C_2$ and $B^*C_3$, since they behave in a similar way to $B^*C_1$.

\section{Final result}

We have seen that the terms $B^{*}C_{n}$ with $n=1,2,3$ give a null contribution to the emission rate while $A_{1}$ and $A_{2}$ contribute as:
\begin{equation}
\frac{d\Gamma}{dt}=
\sum_{\lambda}\int d\Omega_{k}\frac{\gamma}{\hbar^2}\left[R_{11}+2\textrm{Re}\left(R_{12}\right)+R_{22}\right]
\end{equation}
with $R_{11}$, $R_{12}$ and $R_{22}$ given respectively by Eqs.~\eqref{eq:6R1def}, \eqref{eq:6R2def} and \eqref{eq:6R3def}.
The term in the square brackets can be rewritten as:
\[
R_{11}+2\textrm{Re}\left(R_{12}\right)+R_{22}=R_{11}+R_{12}+R_{12}^{*}+R_{22}=
\]
\begin{equation}
=\sum_{f}\sum_{\ell}\left|\sum_{n}\frac{\left\langle f\left|R_{k}\right|n\right\rangle \left\langle n\left|N_{\ell}\right|i\right\rangle }{\left[i\left(\triangle_{fn}+\omega_{k}\right)-\Gamma_{n}\right]}-\frac{\left\langle f\left|N_{\ell}\right|n\right\rangle \left\langle n\left|R_{k}\right|i\right\rangle }{\left[i\left(\triangle_{ni}+\omega_{k}\right)+\Gamma_{n}\right]}\right|^{2}\tilde{f}\left(\triangle_{fi}+\omega_{k}\right).
\end{equation}
Then the formula for the emission rate becomes:
\begin{equation}\label{final}
\frac{d\Gamma}{dt}=\sum_{\lambda}\int d\Omega_{k}\frac{\gamma}{\hbar^2}\sum_{f}\sum_{\ell}\left|\sum_{n}\frac{\left\langle f\left|R_{k}\right|n\right\rangle \left\langle n\left|N_{\ell}\right|i\right\rangle }{\left[i\left(\triangle_{fn}+\omega_{k}\right)-\Gamma_{n}\right]}-\frac{\left\langle f\left|N_{\ell}\right|n\right\rangle \left\langle n\left|R_{k}\right|i\right\rangle }{\left[i\left(\triangle_{ni}+\omega_{k}\right)+\Gamma_{n}\right]}\right|^{2}\tilde{f}\left(\triangle_{fi}+\omega_{k}\right).
\end{equation}
As expected, in Eq.~\eqref{final} unphysical terms proportional to $\tilde{f}(0)$ are not present. This result agrees with that of ref.~\cite{ar}, with two differences. First, it holds for any situation where a quantum system interacts with an external noise; second and more importantly, its rigorous derivation clarifies a long debated issue about the origin of the unphysical terms in the emission rate formula and how to eliminate them.

\section*{Acknowledgements}
This paper is dedicated to S.L. Adler, with whom the authors enjoyed (and still enjoy) many fascinating discussions over this topic and over quantum foundations in general.
Both authors wish to express their gratitude also to D.-A. Deckert for many enjoyable discussions about electromagnetism.
The authors acknowledge support from NANOQUESTFIT, the John Templeton foundation (grant 39530), the COST Action MP1006 and INFN, Italy. 

\section*{Appendix A: Derivation of Eqs.~(\ref{eq:6A1})-(\ref{eq:6C3})}

In this appendix we derive Eqs.~(\ref{eq:6A1})-(\ref{eq:6C3}) using the standard perturbative approach. We want to compute perturbatively the transition probability given in Eq.~(\ref{eq:6Pfinew}):
\begin{equation}
P_{fi}=\mathbb{E}[\left|\left\langle f;\mathbf{k},\lambda\left|U\left(t,t_{i}\right)\right|i;\Omega\right\rangle \right|^{2}]=\mathbb{E}[\left|\left\langle f;\mathbf{k},\lambda\left|U_{I}\left(t,t_{i}\right)\right|i;\Omega\right\rangle \right|^{2}]
\end{equation} 
where $U_{I}\left(t,t_{i}\right)=e^{\frac{i}{\hbar}H_{0}t}U\left(t,t_{i}\right)e^{-\frac{i}{\hbar}H_{0}t_{i}}$ is the time evolution operator in the interaction picture. We recall that here $|i\rangle$ and $|f\rangle$ represent the initial and final state of the system while $|\Omega\rangle$ and $|\mathbf{k},\lambda\rangle$ represent, respectively, the vacuum state of the electromagnetic field and the state with one photon with wave vector $\mathbf{k}$ and polarization $\lambda$. We expand $U_{I}\left(t,t_{i}\right)$ according to the Dyson series~\cite{Sakurai}:
\begin{equation}\label{eq:6dyson}
U_{I}\left(t,t_{i}\right)=1+\sum_{n=1}^{\infty}\left(\frac{-i}{\hbar}\right)^{n}\int_{t_{i}}^{t}dt_{1}\int_{t_{i}}^{t_{1}}dt_{2}...\int_{t_{i}}^{t_{n-1}}dt_{n}H_{1I}\left(t_{1}\right)H_{1I}\left(t_{2}\right)...H_{1I}\left(t_{n}\right)
\end{equation}
where $H_{1I}(t)=e^{\frac{i}{\hbar}H_{0}t}H_{1}(t)e^{-\frac{i}{\hbar}H_{0}t}$ with $H_1(t)$ defined in Eq.~(\ref{H_1}). For the following calculation, it is convenient to rewrite $H_{1}(t)$, inserting the plane waves expansion for the potential vector ${\bf A}({\bf x})$ as given in Eq.~(\ref{AAA}):
\begin{equation}\label{H_1app}
H_{1}(t)=\int d{\bf k}\sum_{\lambda}R_{k}^{\dagger}\, a_{\mathbf{k},\lambda}+\int d{\bf k}\sum_{\lambda}R_{k}\, a_{\mathbf{k},\lambda}^{\dagger}-\sqrt{\gamma}\hbar\sum_{\ell}N_{\ell}w_{\ell}\left(t\right) 
\end{equation}
with $R_k$ defined in Eq.~(\ref{eq:6R}). The first term in Eq.~(\ref{H_1app}) contributes to the processes where a photon is destroyed, the second term to the processes where a photon is created and the last term to the processes where the number of photons is conserved.

We now focus on computing the transition amplitude:
\begin{equation}\label{eq:6Tfi}
T_{fi}:=\left\langle f;\mathbf{k},\mu\left|U_{I}\left(t,t_{i}\right)\right|i;\Omega\right\rangle.
\end{equation}
We will show that, when the Dyson series Eq.~(\ref{eq:6dyson}) is substituted in Eq.~(\ref{eq:6Tfi}), the lowest order terms correspond to the contributions given in Eqs.~(\ref{eq:6A1})-(\ref{eq:6C3}). We start with the zero order term of the Dyson series, which gives a null contribution:
\begin{equation}\label{eq:zero}
\left\langle f;\mathbf{k},\mu|i;\Omega\right\rangle=0
\end{equation}
since the initial and the final states are orthogonal because they contain different number of photons. The next term is the one of Eq.~(\ref{eq:6dyson}) corresponding to $n=1$:
\begin{equation}\label{eq:one}
-\frac{i}{\hbar} \int_{t_{i}}^{t}dt_{1} \left\langle f;\mathbf{k},\mu\left|H_{1I}\left(t_{1}\right)\right|i;\Omega\right\rangle=-\frac{i}{\hbar} \int_{t_{i}}^{t}dt_{1}\, e^{\frac{i}{\hbar}(E_f+\hbar \omega_k-E_i)t_1} \left\langle f;\mathbf{k},\mu\left|H_{1}\left(t_{1}\right)\right|i;\Omega\right\rangle.
\end{equation}
Only the second term of $H_1(t_1)$ gives a contribution in the matrix element $\left\langle f;\mathbf{k},\mu\left|H_{1}\left(t_{1}\right)\right|i;\Omega\right\rangle$, since the other two terms lead to an initial and final state with different numbers of photons. Then we are left with:
\begin{equation}\label{eq:one2}
-\frac{i}{\hbar}\int_{t_{i}}^{t}dt_{1}\left\langle f;\mathbf{k},\mu\left|H_{1I}\left(t_{1}\right)\right|i;\Omega\right\rangle =-\frac{i}{\hbar}\int_{t_{i}}^{t}dt_{1}\, e^{\frac{i}{\hbar}(E_{f}+\hbar\omega_{k}-E_{i})t_{1}}\left\langle f\left|R_{k}\right|i\right\rangle 
\end{equation}
which is exactly the term  ``B" of Eq.~(\ref{eq:6B}).

We then proceed to studying the terms of the Dyson expansion corresponding to $n=2$ (which are the ones corresponding to two-vertex diagrams): 
\begin{equation}\label{eq:second}
\left(\frac{-i}{\hbar}\right)^{2}\int_{t_{i}}^{t}dt_{1}\int_{t_{i}}^{t_{1}}dt_{2}\left\langle f;\mathbf{k},\mu\left|H_{1I}\left(t_{1}\right)H_{1I}\left(t_{2}\right)\right|i;\Omega\right\rangle=
\end{equation}
\begin{eqnarray}
&=&\left(\frac{-i}{\hbar}\right)^{2}\sum_{n}\int_{t_{i}}^{t}dt_{1}\int_{t_{i}}^{t_{1}}dt_{2}\left\langle f;\mathbf{k},\mu\left|H_{1I}\left(t_{1}\right)\right|n;\Omega\right\rangle \left\langle n;\Omega\left|H_{1I}\left(t_{2}\right)\right|i;\Omega\right\rangle+\nonumber\\
\nonumber\\
&+&\left(\frac{-i}{\hbar}\right)^{2}\sum_{n}\int d{\bf k}'\sum_{\mu'}\int_{t_{i}}^{t}dt_{1}\int_{t_{i}}^{t_{1}}dt_{2}\left\langle f;\mathbf{k},\mu\left|H_{1I}\left(t_{1}\right)\right|n;\mathbf{k}',\mu'\right\rangle \left\langle n;\mathbf{k}',\mu'\left|H_{1I}\left(t_{2}\right)\right|i;\Omega\right\rangle. \nonumber
\end{eqnarray}
In the above equation we inserted the completeness over the system states $|n\rangle$ and the one over the Fock space of photons. The latter involves only completeness on states with zero or one photon since for any state involving two or more photons the matrix elements in Eq.~(\ref{eq:second}) are null. The first term of Eq.~(\ref{eq:second}) is the term $A_1$ of  Eq.~(\ref{eq:6A1}), in fact:
\begin{eqnarray}
&&\left(\frac{-i}{\hbar}\right)^{2}\sum_{n}\int_{t_{i}}^{t}dt_{1}\int_{t_{i}}^{t_{1}}dt_{2}\left\langle f;\mathbf{k},\mu\left|H_{1I}\left(t_{1}\right)\right|n;\Omega\right\rangle \left\langle n;\Omega\left|H_{1I}\left(t_{2}\right)\right|i;\Omega\right\rangle =\\
\nonumber&&\\
&&=\left(\frac{-i}{\hbar}\right)^{2}\sum_{n}\int_{t_{i}}^{t}dt_{1}\int_{t_{i}}^{t_{1}}dt_{2}e^{\frac{i}{\hbar}(E_{f}+\hbar\omega_{k}-E_{n})t_{1}}e^{\frac{i}{\hbar}(E_{n}-E_{i})t_{2}}\times   \nonumber\\
\nonumber&&\\
&&\times \left\langle f;\mathbf{k},\mu\left|\int d{\bf k}'\sum_{\lambda}R_{k'}\, a_{\mathbf{k}',\lambda}^{\dagger}\right|n;\Omega\right\rangle \left\langle n;\Omega\left|(-\sqrt{\gamma}\hbar)\sum_{\ell}N_{\ell}\,w_{\ell}\left(t_{2}\right)\right|i\right\rangle =  \nonumber\\
\nonumber&&\\
&&=\left(\frac{-i}{\hbar}\right)^{2}\left(-\sqrt{\gamma}\hbar\right)\sum_{\ell}\sum_{n}\int_{t_{i}}^{t}dt_{1}\int_{t_{i}}^{t_{1}}dt_{2}e^{\frac{i}{\hbar}(E_{f}+\hbar\omega_{k}-E_{n})t_{1}}e^{\frac{i}{\hbar}(E_{n}-E_{i})t_{2}}w_{\ell}\left(t_{2}\right)\left\langle f\left|R_{k}\right|n\right\rangle \left\langle n\left|N_{\ell}\right|i\right\rangle.\nonumber
\end{eqnarray}
In the same way, we can show that the term $A_2$ introduced in Eq.~(\ref{eq:6A2}) is given by the second term of  Eq.~(\ref{eq:second}):
\begin{eqnarray}
&&\left(\frac{-i}{\hbar}\right)^{2}\sum_{n}\int d{\bf k}'\sum_{\mu'}\int_{t_{i}}^{t}dt_{1}\int_{t_{i}}^{t_{1}}dt_{2}\left\langle f;\mathbf{k},\mu\left|H_{1I}\left(t_{1}\right)\right|n;\mathbf{k}',\mu'\right\rangle \left\langle n;\mathbf{k}',\mu'\left|H_{1I}\left(t_{2}\right)\right|i;\Omega\right\rangle =\\
\nonumber&&\\
&&=\left(\frac{-i}{\hbar}\right)^{2}\sum_{n}\int d{\bf k}'\sum_{\mu'}\int_{t_{i}}^{t}dt_{1}\int_{t_{i}}^{t_{1}}dt_{2}e^{\frac{i}{\hbar}(E_{f}+\hbar\omega_{k}-E_{n}-\hbar\omega_{k'})t_{1}}e^{\frac{i}{\hbar}(E_{n}+\hbar\omega_{k'}-E_{i})t_{2}}\times   \nonumber\\
\nonumber&&\\
&&\times \left\langle f;\mathbf{k},\mu\left|H_{1}\left(t_{1}\right)\right|n;\mathbf{k}',\mu'\right\rangle \left\langle n;\mathbf{k}',\mu'\left|H_{1}\left(t_{2}\right)\right|i;\Omega\right\rangle=   \nonumber\\
\nonumber&&\\
&&=\left(\frac{-i}{\hbar}\right)^{2}\sum_{n}\int d{\bf k}'\sum_{\mu'}\int_{t_{i}}^{t}dt_{1}\int_{t_{i}}^{t_{1}}dt_{2}e^{\frac{i}{\hbar}(E_{f}+\hbar\omega_{k}-E_{n}-\hbar\omega_{k'})t_{1}}e^{\frac{i}{\hbar}(E_{n}+\hbar\omega_{k'}-E_{i})t_{2}}\times \nonumber\\
\nonumber&&\\
&&\times\left\langle f;\mathbf{k},\mu\left|(-\sqrt{\gamma}\hbar)\sum_{\ell}N_{\ell}\,w_{\ell}\left(t_{1}\right)\right|n;\mathbf{k}',\mu'\right\rangle \left\langle n;\mathbf{k}',\mu'\left|\int d{\bf k}''\sum_{\lambda}R_{k''}\, a_{\mathbf{k}'',\lambda}^{\dagger}\right|i;\Omega\right\rangle\nonumber\\
\nonumber&&\\
&&=\left(\frac{-i}{\hbar}\right)^{2}\left(-\sqrt{\gamma}\hbar\right)\sum_{\ell}\sum_{n}\int_{t_{i}}^{t}dt_{1}\int_{t_{i}}^{t_{1}}dt_{2}e^{\frac{i}{\hbar}(E_{f}-E_{n})t_{1}}e^{\frac{i}{\hbar}(E_{n}+\hbar\omega_{k}-E_{i})t_{2}}w_{\ell}\left(t_{1}\right)\left\langle f\left|N_{\ell}\right|n\right\rangle \left\langle n\left|R_{k}\right|i\right\rangle.\nonumber 
\end{eqnarray}
In a similar way, one can show that the terms $C_1$, $C_2$ and $C_3$ of Eqs.~(\ref{eq:6C1})-(\ref{eq:6C3}), which are represented by diagrams containing three vertices, can be obtain from the term of the Dyson expansion  Eq.~(\ref{eq:6dyson}) corresponding to $n=3$.

\section*{Appendix B: Calculation of the natural broadening for an harmonic oscillator}
In this appendix we compute the natural broadening for an harmonic oscillator. The starting point is the equation for the imaginary part of the energy shift which, in the case of non relativistic electromagnetic interactions, is given in~\cite{Sakurai2} (page 67):
\begin{equation}\label{deltaim1}
\triangle E_{i}=-\pi\sum_{\lambda}\int d\mathbf{k}\sum_{n}\left|\left\langle \mathbf{k},\lambda;n\left|H_{int}\right|\Omega;i\right\rangle \right|^{2}\delta\left(E_{i}-E_{n}-\hbar\omega_{k}\right),
\end{equation}
where $|i\rangle$ and $|n\rangle$ are the eigenstates of the harmonic oscillator Hamiltonian $H_0$ with eigenvalues $E_i$ and $E_n$ and $|\Omega\rangle$ and $|\mathbf{k},\lambda\rangle$ represent, respectively, the vacuum state of the electromagnetic field and the state with one photon with wave vector $\mathbf{k}$ and polarization $\lambda$.  
 We work in dipole approximation, so that the interaction Hamiltonian becomes:
\begin{equation}
H_{int}=-\frac{e}{m}\mathbf{A}\cdot\mathbf{p}=-\frac{e}{m}\sum_{\lambda}\int d\mathbf{k}\,\alpha_{k}\left(a_{\mathbf{k},\lambda}+a_{\mathbf{k},\lambda}^{\dagger}\right)\left(\epsilon_{\mathbf{k},\lambda}\cdot\mathbf{p}\right).
\end{equation}
Then Eq.~\eqref{deltaim1} becomes:
\begin{eqnarray}\label{imshift}
\triangle E_{i}&=&-\left(\frac{e}{m}\right)^{2}\pi\sum_{\lambda}\int d\mathbf{k}\alpha_{k}^{2}\sum_{n}\left|\left\langle n\left|\epsilon_{\mathbf{k},\lambda}\cdot\mathbf{p}\right|i\right\rangle \right|^{2}\delta\left(E_{i}-E_{n}-\hbar\omega_{k}\right)\\
 &=&-\left(\frac{e}{m}\right)^{2}\pi\sum_{\lambda}\int d\mathbf{k}\alpha_{k}^{2}\sum_{n}\left\langle i\left|\epsilon_{\mathbf{k},\lambda}\cdot\mathbf{p}\right|n\right\rangle \left\langle n\left|\delta\left(E_{i}-H_{0}-\hbar\omega_{k}\right)\epsilon_{\mathbf{k},\lambda}\cdot\mathbf{p}\right|i\right\rangle \nonumber\\
 &=&-\left(\frac{e}{m}\right)^{2}\pi\sum_{\lambda}\int d\mathbf{k}\alpha_{k}^{2}\left\langle i\left|\epsilon_{\mathbf{k},\lambda}\cdot\mathbf{p}\,\delta\left(E_{i}-H_{0}-\hbar\omega_{k}\right)\epsilon_{\mathbf{k},\lambda}\cdot\mathbf{p}\right|i\right\rangle \nonumber 
\end{eqnarray}
where in the third line we used the completeness over the states $|n\rangle$. Note that $H_0$ is an operator, therefore the delta function cannot be brought out of the scalar product.

It is now convenient to write the momentum operator $\mathbf{p}$ in terms of the raising and lowering operators $b_{j}^{\dagger}$ and $b_{j}$, where $j$ labels the three spatial components. Then we get:
\begin{equation}
\epsilon_{\mathbf{k},\lambda}\cdot\mathbf{p}=\sum_{j=1}^{3}\epsilon_{\mathbf{k},\lambda}^{j}i\sqrt{\frac{m\omega_{0}\hbar}{2}}\left[b_{j}^{\dagger}-b_{j}\right]
\end{equation}
Substituting this expression in Eq.~\eqref{imshift} and using the relation: 
\begin{equation}
\sum_{\lambda}\int d\Omega_{\mathbf{k}}\epsilon_{\mathbf{k}\lambda}^{j}\epsilon_{\mathbf{k}\lambda}^{j'}=\frac{8}{3}\pi\delta_{jj'},
\end{equation}
we obtain 
\begin{equation}\label{boboapp1}
\triangle E_{i}=\frac{4e^{2}\pi^{2}\omega_0 \hbar}{3m}\int dk\, k^{2}\alpha_{k}^{2}\sum_{j=1}^{3}\left\langle i\left|\left(b_{j}^{\dagger}-b_{j}\right)\delta\left(E_{i}-H_{0}-\hbar\omega_{k}\right)\left(b_{j}^{\dagger}-b_{j}\right)\right|i\right\rangle.
\end{equation}
It is straightforward to show that the matrix element is equal to:
\begin{equation}
\left\langle i\left|\left(b_{j}^{\dagger}-b_{j}\right)\delta\left(E_{i}-H_{0}-\hbar\omega_{k}\right)\left(b_{j}^{\dagger}-b_{j}\right)\right|i\right\rangle =-\left(i_{j}+1\right)\delta\left(\hbar\omega_{0}+\hbar\omega_{k}\right)-i_{j}\delta\left(\hbar\omega_{0}-\hbar\omega_{k}\right),
\end{equation}
where $i_{j}$ with $j=1,2,3$ is one of the three quantum numbers which identify the initial energy state, i.e. $E_{i}=\hbar\omega_{0}\left(\frac{3}{2}+i_{1}+i_{2}+i_{3}\right)$. Therefore Eq.~\eqref{boboapp1} becomes:
\begin{eqnarray}\label{basta}
\triangle E_{i}&=&\frac{4e^{2}\pi^{2}\omega_0 \hbar}{3mc^{3}}\int d\omega_{k}\,\omega_{k}^{2}\left(\frac{\hbar}{2\varepsilon_{0}\omega_{k}\left(2\pi\right)^{3}}\right)\times \\
\nonumber\\
&\times&\sum_{j}\left[-\left(i_{j}+1\right)\delta\left(\hbar\omega_{0}+\hbar\omega_{k}\right)-i_{j}\delta\left(\hbar\omega_{0}-\hbar\omega_{k}\right)\right],\nonumber
\end{eqnarray}
where we used $\alpha_{k}^{2}=\frac{\hbar}{2\varepsilon_{0}\omega_{k}\left(2\pi\right)^{3}}$ and we performed the change of variable $k\rightarrow\omega_{k}=kc$. The first delta function never contributes, then Eq.~\eqref{basta} becomes:
\begin{equation}
\triangle E_{i}=-\hbar\left(\frac{\beta\omega_{0}^{2}}{2m}\right)\sum_{j}i_{j}\, ,
\end{equation}
where we introduced $\beta=\frac{e^{2}}{6c^{3}\pi\varepsilon_{0}}$. The decay is proportional to $-\frac{\triangle E_{i}}{\hbar}=\left(\frac{\beta\omega_{0}^{2}}{2m}\right)\sum_{j}i_{j}$, which is proportional to the decay rate $\Lambda=\frac{\omega_0^2 \beta}{2m}$ found in~\cite{basdon,dirk}.

\end{document}